\documentclass{evn2004}
\usepackage{txfonts}
\usepackage{graphicx}
\def\asec{\ifmmode ^{\prime\prime}\else$^{\prime\prime}$\fi}
\def\etal{{et\,al. }}
\def\Ms{\mbox{\,M$_{\odot}$}}

\def\msun{\hbox{M$_{\odot}$}}

\def\msunyr{\mbox{\,${\rm M_{\odot}\, yr^{-1}}$}}
\def\mdot{\dot M}
\def\Mdot{\dot M}

\def\degs{\ifmmode ^{\circ}\else$^{\circ}$\fi}
\def\amin{\ifmmode ^{\prime}\else$^{\prime}$\fi}
\def\asec{\ifmmode ^{\prime\prime}\else$^{\prime\prime}$\fi}
\def\fdg{\hbox{$.\!\!^\circ$}}          

\def\degs{\ifmmode ^{\circ}\else$^{\circ}$\fi}
\def\amin{\ifmmode ^{\prime}\else$^{\prime}$\fi}

\def\EE#1{\times 10^{#1}}

\def\cm{\mbox{\,cm}}

\def\cm3{\mbox{\,cm$^{-3}$}}
\def\kms{\mbox{\,km~s$^{-1}$}}

\def\ergs{\mbox{\,erg~s$^{-1}$}}

\def\kms{\mbox{\,km s$^{-1}$}}

\def\lsim{\!\!\!\phantom{\le}\smash{\buildrel{}\over
 {\lower2.5dd\hbox{$\buildrel{\lower2dd\hbox{$\displaystyle<$}}\over
                                 \sim$}}}\,\,}
\def\gsim{\!\!\!\phantom{\ge}\smash{\buildrel{}\over
{\lower2.5dd\hbox{$\buildrel{\lower2dd\hbox{$\displaystyle>$}}\over
                               \sim$}}}\,\,}

\begin{document}
\setcounter{page}{225}

   \title{VLBI observations of young Type II supernovae}

   \author{Miguel A.\ P\'erez-Torres
          \inst{1}
          \and A.\ Alberdi\inst{1}
          \and J.M.\ Marcaide\inst{2}
         }

   \institute{Instituto de Astrof\'{\i}sica de Andaluc\'{\i}a, CSIC, Apdo.
           Correos 3004, E-18080 Granada, Spain \\ 
           email: torres@iaa.es, antxon@iaa.es
        \and 
           Departamento de Astronom\'{\i}a y 
           Astrof\'{\i}sica, Universidad de Valencia,
           46100 Burjassot, Spain \\
           email: J.M.Marcaide@uv.es
           }

\abstract{
We give an overview of circumstellar interaction 
in young Type II supernovae, as seen through the eyes of
very-long-baseline interferometry (VLBI) observations.
The resolution attained by such observations ($\lsim 1$~mas)
is a powerful tool to probe the 
interaction that takes place after a supernova goes off. 
The direct imaging of a supernova permits, in principle, 
to estimate the deceleration of its expansion, and
to obtain information on the ejecta and circumstellar density
profiles, as well as estimates of the magnetic field intensity
and relativistic particle energy density in the supernova.
Unfortunately, only a handful of radio supernovae are close and bright
enough as to permit their study with VLBI.
We present results from our high-resolution observations of the
nearby Type II radio supernovae SN~1986J and SN~2001gd. 
}

   \maketitle
%

\section{Introduction}

Radio supernovae are mostly associated with core-collapse supernovae
(Type II and Ib/c). 
Here, we will discuss only Type II radio supernovae, although 
most results apply as well to Type Ib/c radio supernovae.
The progenitors of Type II supernovae are surrounded by 
a high-density wind, $n \approx 3\EE{7}
\Mdot_{-5}\,r_{15}\,v^{-1}_1\,\cm3$, where
$\Mdot_{-5}$ is the mass loss rate of the presupernova
wind in units of 10$^{-5}\,\msunyr$, $v_1$ is the presupernova
wind velocity in units of $10\,\kms$, and 
$r_{15}$ is the supernova shock radius in units of $10^{15}$~cm.
In the standard interaction model for radio supernovae 
(Chevalier~\cite{chevalier82}, Nadyozhin~\cite{nadyozhin85}),
a forward shock at velocities thar largely exceed $\sim 10000\,\kms$
propagates into the circumstellar medium (CSM) 
when the blastwave reaches the dense, ionized, slowly expanding wind. 
In addition, a reverse shock moves back into the stellar envelope
at speeds of $\sim\,1000\,\kms$ relative to the expanding ejecta.
As a result, a high-energy-density shell is formed.

The brightness temperature inferred from 
VLBI images of SNe is very large, indicating that the emission
is of non-thermal origin, namely synchrotron radiation from relativistic
electrons ($N_E = N_0\,E^{-\gamma}$) in the high-energy-density shell. 
As it expands, a radio supernova quickly increases its radio brightness 
with time, due to the increasingly smaller electron column density 
in the line of sight.  
When the optical depth at cm-wavelengths has dropped to about unity, the 
supernova reaches its maximum of emission,
after which the emission monotonically decreases. 
If synchrotron self-absorption effects are negligible, 
then the only relevant source of absorption is 
free-free absorption by thermal electrons in the CSM,
and the radio luminosity evolves as
$L_\nu \propto R^2\,\Delta R\,N_0\,B^\frac{\gamma + 1}{2}\,
    \nu^{-\frac{\gamma - 1}{2}}\, e^{-\tau_{\rm \nu, ff}}$
($\tau_{\rm ff, \nu} \propto \nu^{-2.1}$).
Thus, a radio supernova is first seen at high
frequencies and, as times goes, it becomes visible at increasingly
lower frequencies. 
We see that the flux density monitoring of radio supernovae may yield important
information on such fundamental physical parameters as magnetic field 
intensity and relativistic particle energy density in the supernova
shell, as well as information on the relevant absorption and cooling 
mechanisms. 
Since the radio peaks of SNe are usually of the order of a few milliJansky,  
only bright, nearby radio supernovae are good VLBI targets.

The density of the supernova ejecta can be approximated by a
steep power-law profile ($\rho_{ej} \propto r^{-n}$).
If the mass loss parameters ($\Mdot, v_w$) stay 
approximately constant up to the
explosion, the CSM density profile is then 
$\rho_{csm} \propto r^{-2}$.
The dense shell formed at shock breakout 
decelerates as it interacts with the CSM, and
it can be shown that the expansion of the shell radius 
follows a power-law with time:
$R \propto t^m; m=(n-3)/(n-2)$. 
More generally, $\rho_{\rm csm} \propto r^{-s}$,
leading to $m = (n-3)/(n-s)$.
Therefore, by carrying several VLBI observations of the same supernova, 
we may be able to infer the value of its deceleration index, $m$, and 
indirectly the values of $n$ and $s$ that characterize the ejecta and 
CSM density profiles.

   \begin{figure*}
   \centering
   \includegraphics[width=\textwidth]{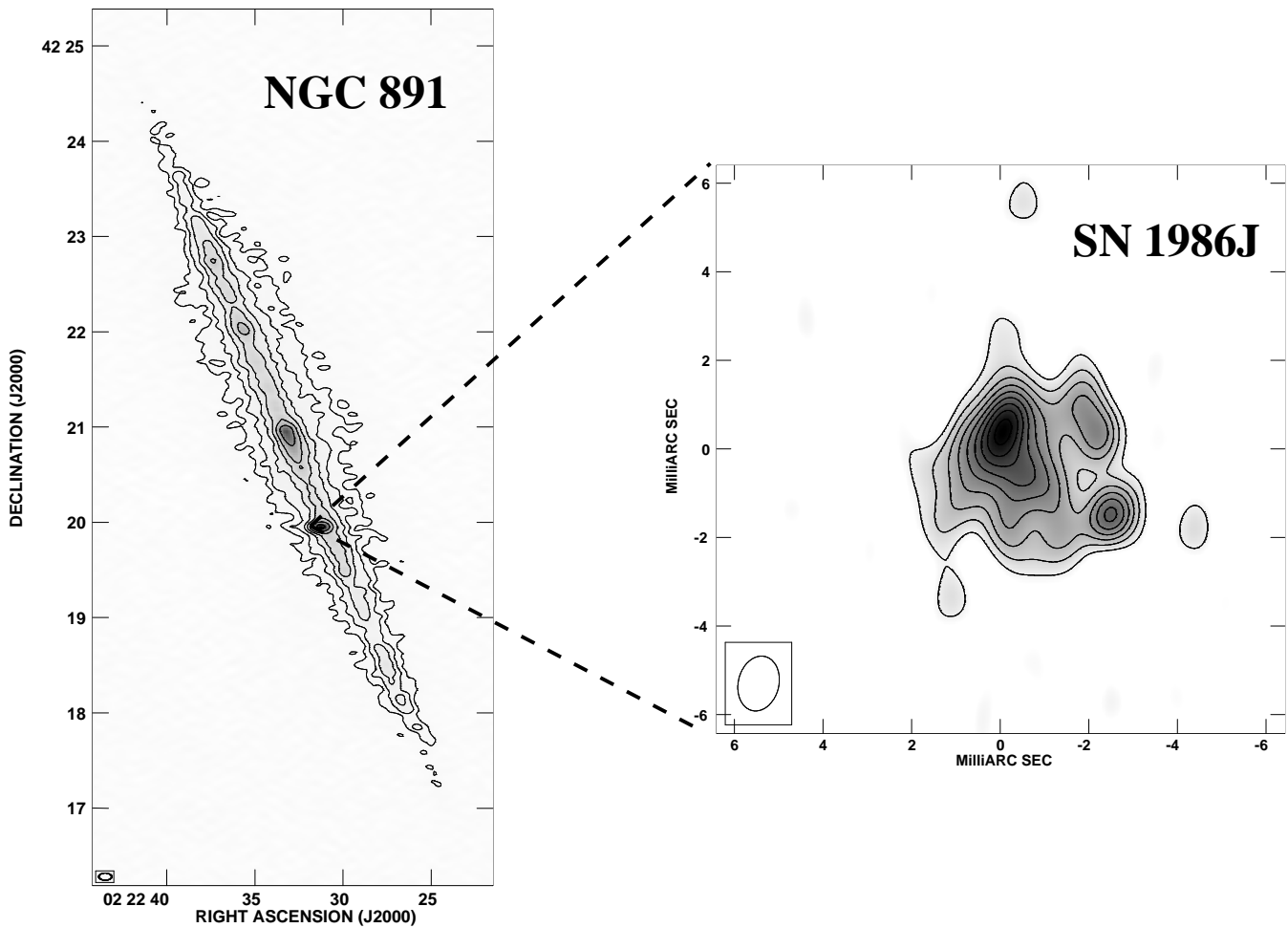}
      \caption{
{\em Left}: 5~GHz VLA map of the galaxy NGC~891 and its supernova SN\,1986J
        made on February 21, 1999.
        The contours are (1,2,4,8,...)$\times$\,106\,$\mu$Jy beam$^{-1}$.
        The peak of brightness of the map corresponds to the supernova
        and is $\sim$ 8.2 mJy beam$^{-1}$.
{\em Right}: Global VLBI map of SN\,1986J on February 21, 1999.
       The contours are (3,5,7,9,11,13,15,17,19) $\times$ 56 $\mu$Jy beam$^{-1}$,
        the off-source rms.
        The peak of brightness of the map is 1.13 mJy beam$^{-1}$ and
        the restoring beam (bottom left in the image) is $1.3 \times 0.9$ mas$^2$
        at a position angle of -13$\fdg$4.
        In both panels, north is up and east is left.
              }
         \label{fig,sn1986j}
   \end{figure*}
%

\section{SN~1986J in NGC~891}
\label{sec,sn1986j}

SN\,1986J in NGC\,891 is one of the most radio luminous SNe ever discovered.
The precise date of its explosion is not known, but
on the basis of the available radio and optical data
SN\,1986J was estimated to have exploded around the end of 1982, 
or the beginning of 1983  
(e.g., Weiler, Panagia \& Sramek 1990).
Based upon its large radio luminosity, 
Weiler et al.~\cite{weiler90} suggested that
the progenitor star was probably a red giant with
a main-sequence mass of (20 -- 30)\,\Ms\, that had lost material
rapidly ($\mdot\,\ga2\,\times\,10^{-4}\,\msunyr$) in a dense stellar
wind. 
VLBI observations made at 8.4\,GHz at 
the end of 1988 by Bartel et al.~(\cite{bartel91}, hereafter B91) 
showed that the radio
structure of SN1986J had the form of a shell, in agreement with 
expectations from the standard interaction model, with 
the minimum of emission located approximately at its center.
Those authors claimed the existence of several protrusions at
distances of twice the shell radius, and with apparent
expansion velocities as high as $\sim 15000 \kms$.
Since these protrusions were twice as far as the mean radius of the
shell, it then follows that the main bulk of the shell expanded at roughly
$7500 \kms$.
Such protrusions have been successfully invoked by Chugai (\cite{chugai93}) to explain
the coexistence of velocities smaller than 1000$\kms$ implied from the
observed narrow optical lines (Leibundgut et al.~\cite{leibundgut91}),
and the large velocities indicated from the VLBI measurements.

We used archival VLA and global VLBI observations
of supernova 1986J at 5\,GHz, taken about 16 yr after the
explosion, to obtain the images shown in Fig.~\ref{fig,sn1986j}.
The right panel corresponds to the 5\,GHz VLBI image of SN1986J.
It shows a highly distorted shell of radio emission, indicative of a
strong deformation of the shock front.
The apparent anisotropic brightness distribution
is very suggestive of the forward shock
colliding with a clumpy, or filamentary wind.
Note that there are several ``protrusions'' outside the shell,
though just above three times the noise level and at different position
angles from those previously reported by B91.
Therefore, these protrusions could not be real, but must be just artifacts of the image
reconstruction procedure.
If this is the case, the disappearance
of the protrusions seen in the previous VLBI observations (B91)
would imply a change in the density profile of the circumstellar wind.

At a distance of 9.6 Mpc (Tully 1998), 1 mas corresponds to
a linear size of $\sim1.4\, \times\, 10^{17}\,{\rm cm}\,\approx0.05\,{\rm pc}$.
Based on 8.4 GHz VLBI observations,
B91  found an angular size  of $\sim$3.7 mas
for the shell of SN\,1986J, 5.74 yr
after its explosion (assuming it took place on 1983.0).
The corresponding mean linear velocity would then be
$\sim$14700\,km\,s$^{-1}$ for the first 5.74 yr.
However, this velocity applies only to the protrusions found by B91, not
to the shell.
Indeed, the velocities reported in B91
were calculated for the protrusions, and assuming that these
originated in the centre at the time of the explosion.
These authors also pointed out that the protrusions extended
from the centre to twice the radius of the shell, i.e.,
the protrusions were {\it outside} the shell.
Therefore, a value of $\sim$1.85 mas
for the angular size of the shell of SN\,1986J
at epoch 1988.74 is indicated, and a mean linear velocity of the
radio shell of $\sim$7400\,km\,s$^{-1}$ is more appropriate
to characterize the first 5.74 yr of the expansion of SN\,1986J,
as has been previously noticed by Chevalier (\cite{chevalier98}) and Houck
et al.~(\cite{houck98}).

   \begin{figure*}
   \centering
   \includegraphics[width=\textwidth]{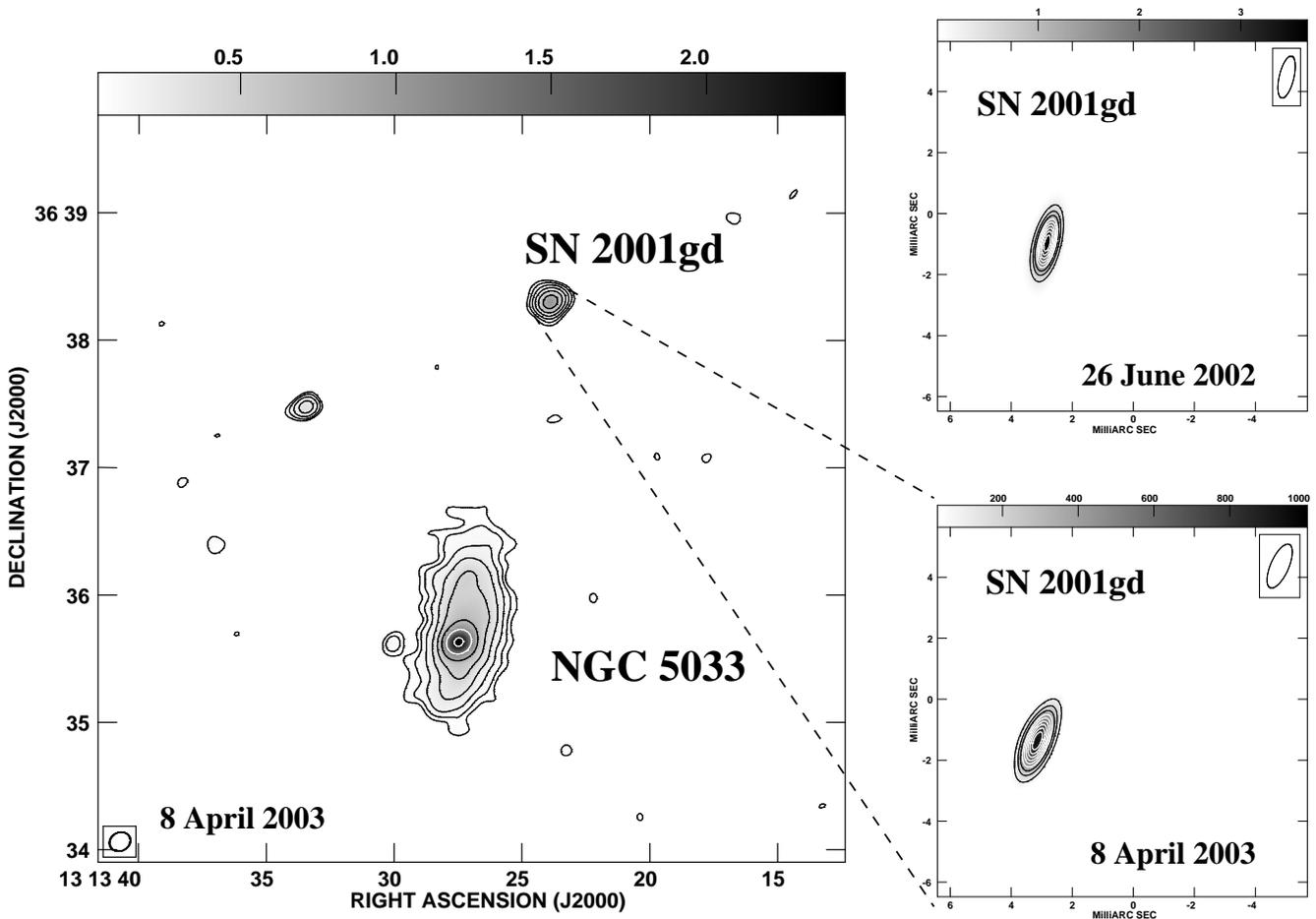}
      \caption{
{\em Left}: 8.4~GHz VLA  map of the galaxy NGC\,5033 and its supernova SN\,2001gd, 
from observations on 8 April 2003.
The lowest contour is at 48$\mu$Jy beam$^{-1}$, corresponding to 
three times the off-source rms.
The peak of brightness of the map corresponds to the nucleus of NGC\,5033
and is 2.44 mJy beam$^{-1}$.
The supernova is the bright point-like source northwards of the
nucleus of NGC\,5033.
{\em  Right}:
(Top) VLBI map of SN\,2001gd on 26 June 2002.
Contours are drawn at 10, 20, 30, 40, 50, 60, 70, 80, and 90\% of the
peak of brightness of 3.64 mJy beam$^{-1}$.
(Bottom) VLBI map of SN\,2001gd on 8 April 2003.
Contours are drawn at 10, 20, 30, 40, 50, 60, 70, 80, and 90\% of the
peak of brightness of 947$\mu$Jy beam$^{-1}$.
In all panels, north is up and east is left.
              }
         \label{fig,sn2001gd}
   \end{figure*}
%

The angular size of the shell of SN~1986J on 21 February 1999
is $\sim$4.7 mas, 
equivalent to $\approx6.8\,\times\, 10^{17}\,$\,cm.
Combining this angular size measurement 
with that obtained by B91 for epoch 1988.74
($\theta\,\approx\,1.85$\,mas), 
we obtain a mean angular expansion velocity of the shell
between 29 September 1988 (1988.74) and 21 February 1999 (1999.14) of
$\approx 0.14\,{\rm mas\, yr^{-1}}$, which corresponds to a linear
velocity of $\sim6300\,{\rm km\,s^{-1}}$.
If we assume that SN\,1986J freely expanded for the
first 5.74 yr of its life  ($r\,\propto\,t^m, m=1$) and
then started to decelerate, 
the expansion between the two epochs of VLBI observations
is characterized by $m = 0.90 \pm 0.06$.  
This mild deceleration contrasts with the case
of several other supernovae, for which 
a strong deceleration has been measured 
(SN1979C:  Marcaide et al.~\cite{marcaide02};
SN1987A: Staveley-Smith et al.~\cite{staveley93}, Gaensler et al.~\cite{gaensler97};
SN1993J: Marcaide et al.~\cite{marcaide97}, Bartel et al.~\cite{bartel00}). 

For a standard presupernova wind velocity, $v_w$=10 \kms,
the linear size of SN\,1986J at epoch $t$=16.14 yr implies that
we are sampling the progenitor wind about 11000 yr prior to
its explosion.
Since the mass loss rate of SN\,1986J seems to have been
$\ga2\,\times\, 10^{-4}\,\msunyr$ (Weiler, Panagia \& Sramek 1990), 
the swept-up mass must have been ${\rm M_{sw}}\,\approx2.2\Ms$
for a standard density profile of the progenitor wind
($\rho_{\rm cs} \propto r^{-2}$).
Since the expansion of the supernova has not decelerated significantly
between the two VLBI observations, it follows that the swept-up
mass by the shock front must be much less than the mass
of the ejected hydrogen-rich envelope, ${\rm M_{env}}$,
as otherwise we should have observed a much stronger deceleration.
In fact, momentum conservation implies that
${\rm M_{env}}\, \ga\, 12\, \Ms$, significantly larger than ${\rm M_{sw}}$.
If the hydrogen-rich mass envelope was as high as 
12\,\Ms, this is a hint that the progenitor of SN\,1986J was probably a 
single, massive Red Super Giant (as previously suggested by Weiler et al. 1990), 
which lost mass rapidly, but managed to keep intact most of
its hydrogen-rich envelope by the time of explosion.
This result contrasts with the cases of SN\,1993J and SN\,1979C, whose
hydrogen-rich envelopes had masses of 0.2--0.4\,\Ms 
(Woosley et al. 1994, Houck \& Fransson 1996) and $\la\,0.9\,\Ms$ 
(Marcaide et al. 2002), and whose progenitor stars were part of binary
systems.

Since the radio emission is of synchrotron origin,
we can estimate a minimum total energy (in magnetic fields,
electrons, and heavy particles) and a minimum magnetic field for SN\,1986J.
If we assume equipartition, then the minimum total energy is (Pacholczyk 1970)
$E_{\rm min}^{\rm (tot)} = c_{13}\, (1 + k)^{4/7}\, \phi^{3/7}\, R^{9/7}\, L^{4/7},$
where $L$ is the radio luminosity of the source,
$R$ is a characteristic size,
$c_{13}$ is a slowly-dependent function of the spectral index, $\alpha$,
$\phi$ is the fraction of the supernova's volume occupied  by the magnetic field and
by the relativistic particles (filling factor), and
$k$ is the ratio of the (total) heavy particle energy to the electron energy.
This ratio depends on the mechanism that generates the relativistic electrons, ranging
from $k \approx 1$ up to $k = m_p/m_ e \approx 2000$, where $m_p$ and $m_e$
are the proton and electron mass, respectively.

Based on snapshot VLA observations of SN\,1986J carried out on 13 June 1999, 
we determined an spectral index of
$\alpha =-0.69 \pm 0.06$ ($S_\nu \propto \nu^{\alpha}$) between 1.6 and 8.5 GHz.
This value is very close to {\bf $\alpha = -0.67$} obtained by B91,
and is a typical value for supernovae that are in the optically thin radio regime.
With $\alpha =-0.69$ and $S_{4.9 \rm GHz} = 7.2$ mJy from our observations, we obtain
a radio luminosity $L_\nu \approx 1.38 \times 10^{37} (D/9.6 {\rm Mpc})^2$ erg\,s$^{-1}$
for the frequency range  $10^7$--$10^{10}$ Hz.
The value of the function $c_{13}$ is approximately $2.39 \times 10^4$.
As the characteristic size for SN\,1986J, we take half the largest
diameter of the shell, $\theta = 2.35\, {\rm mas}$, which corresponds
to a linear size of $R = D\, \cdot\, \theta\, \approx3.4\, \times 10^{17}$cm.
With these values, the minimum total energy is then

\[
 E_{\rm min}^{\rm (tot)} \approx 1.13 \times 10^{48} \,(1 + k)^{4/7}\, 
   \phi_{0.66}^{3/7}\,
   \theta_{2.35}^{9/7}\,   
   D_{9.6}^{17/7}\,  {\rm erg}
\]

\noindent
where $\phi$=0.66$\phi_{0.66}$, 
$\theta$=2.35$\theta_{2.35}$\,mas , and
$D$=9.6$D_{9.6}$\,Mpc.
The value of the magnetic field that yields $E^{\rm (min)}_{\rm tot}$ is
then equal to

\[
 B_{\rm min} \approx 10.6 \, (1 + k)^{2/7}\, 
   \phi_{0.66}^{-2/7}\, 
   \theta_{2.35}^{-6/7}\,   
   D_{9.6}^{-2/7}  {\rm mG}
\]

\noindent
Since $1 \leq k \leq 2000$,
$E_{\rm min}^{\rm (tot)}$ can have values between $\sim 2 \times 10^{48}$
and $\sim 9 \times 10^{49} {\rm erg}$, while the corresponding values of the
magnetic field can lie between $\sim 13$  and
$\sim 93\, {\rm mG}$ (for $\phi = 0.66$ and $D = 9.6 {\rm Mpc}$).
These values of the magnetic field for SN\,1986J are in agreement
with, e.g., those obtained for SN\,1993J at similar radii.
For example, P\'erez-Torres, Alberdi, and Marcaide \cite{mapt01}
showed that $B\approx\,0.58\, (r/3\,\times\,10^{16}\,{\rm cm})^{-1}$\,G,
which results in a value of $\sim\,50$\,mG for a radius of
$3.4\,\times\,10^{17}$\,cm.
Since it is very unlikely that the magnetic field energy density in the
wind is larger than its kinetic energy density, i.e.,
$B^2/8\pi \leq \rho v^2_{\rm w}/2$, we can
then obtain an upper limit for the magnetic field in the circumstellar
wind of SN\,1986J:

\[
 B_{\rm cs} \la (\mdot\,v_w)^{1/2}\,r^{-1}\,\approx\,
                        0.78\,(\mdot_{-4}\,v_{10})^{1/2}\, r_{17}^{-1}\,{\rm mG} 
\]

\noindent
where $\mdot$=$10^{-4}\,\mdot_{-4}\msunyr$, $v_w$=10$v_{10}\,\kms$,
$r$=$10^{17}\,r_{17}\,{\rm cm}$, and
we have assumed a standard wind density profile.
For SN\,1986J, $\mdot = 2\,\times\,10^{-4}\,\msunyr$ and $r\,=\,3.4\,\times\,10^{17}$\,cm,
and we obtain $B_{\rm csm}\, \la \,0.32$\,mG.
These arguments suggest that the magnetic field
in the shell of SN\,1986J is in the range 13--93 mG, or about 40 to 300 times
the magnetic field in the circumstellar wind.
Since a strong shock yields a fourfold increase in the
particle density, the post-shock magnetic field is
also fourfold increased.
Hence, compression alone of the circumstellar wind magnetic field cannot
account for the large magnetic fields existing in SN\,1986J, and
other field amplification mechanisms are needed to be invoked, e.g.,
turbulent amplification (Chevalier 1982; Chevalier \& Blondin 1995).
The same conclusion was reached for the case of SN\,1993J
(Fransson \& Bj\"ornsson 1998, P\'erez-Torres et al.~\cite{mapt01}),
where magnetic field amplification factors $\sim100\,$ were found to be necessary.

\begin{table*}[thb!]
\begin{center}
 \caption[]{Properties of some Type II radio supernovae}
   \label{tab,rsne}
\begin{tabular}{lllll}
\hline\noalign{\smallskip}
                  &  SN~1979C$^1$     &  SN~1986J$^2$  &  SN~2001gd$^3$  & SN~1993J$^4$ \\
\noalign{\smallskip}\hline\noalign{\smallskip}
Distance (Mpc)    &  16.1         &  9.6       &  21.6       & 3.63     \\
Time since 
explosion$^5$ (yr)    &  $\sim 20.1$  &  $\sim 16$ &  $\lsim 2$    & $\sim 8.6$     \\
$(L/L_{\rm SN~1993J})_{6~{\rm cm, peak}}$    
                  &  $\sim 1.6$   &  $\sim 13$ &  $\sim 2$   &  1    \\
Radio brightness structure
                  &Shell (likely) &  Distorted shell & ?     & $\sim$Smooth shell  \\
$\Mdot / 10^{-5} \msunyr$  
                  &  $\sim (12-16)$ &  $\sim 20$ & ?           & $\sim 5$  \\
Deceleration parameter $(m)$   
                  &  $\sim 0.62$  &$\sim 0.90$ & ?           & $\sim 0.82$  \\
$t_{\rm break}$ (years) 
             &  $6 \pm 2$    & Not yet    & $\sim$?           & $\sim 0.5$ \\
Asymmetric expansion?    
                 &  No           & Yes        & ?           & No ($\lsim 5\%$) \\
Circumstellar medium
                 &  ?            & Clumpy     & ?           & Approx. smooth \\
$M_{\rm swept}/\msun$ 
           &  $\sim 1.6$   & $\sim 2.2$ & $\sim$?           & $\sim 0.4$ \\
$M_{\rm env}/\msun$ 
           &  $\sim 0.9$   & $\sim 12$  &  $\sim$?           & $\sim 0.6$ \\
Explosion scenario 
           &  Binary       & Single     &  Binary?           & Binary \\
Magnetic field amplification
           & Turbulent     & Turbulent  &  Turbulent         & Turbulent \\
\noalign{\smallskip}\hline
\end{tabular}
\end{center}

\begin{list}{}{}
\item[]{
$^1$ From Marcaide et al.~(\cite{marcaide02}).
$^2$ From P\'erez-Torres et al.~(\cite{mapt02a}).
$^3$ From P\'erez-Torres et al.~(\cite{mapt04}).
$^4$ From Marcaide et al.~(\cite{marcaide95a}), Marcaide et al.~(\cite{marcaide95b}), 
          Marcaide et al.~(\cite{marcaide97}),
	  P\'erez-Torres et al.~(\cite{mapt01},\cite{mapt02b}). 
$^5$ Time since explosion at which we are discussing the property of the supernova.
}
\end{list}
\end{table*}

\section{SN~2001gd in NGC~5033}
\label{sec,sn01gd}

NGC~2001gd in NGC~5033 was discovered by Nakano et al.~(\cite{nakano01})
on 24.820 November 2001, although its explosion date is uncertain.
The supernova had a visual magnitude of 14.5, and is located
$\sim$3' north-northwest of the nucleus of NGC~5033.
Nakano et al. (\cite{nakano01}) reported the following position
for SN~2001gd: $\alpha$=13h13m23s.89, $\delta$=+36$^\circ$38'17".7.
They also reported that there was no star visible at the
above position on earlier frames taken from 1996 to April 2001.
A spectrum obtained by P. Berlind ten days later (Matheson et al.~\cite{matheson01}),
showed SN~2001gd to be a Type IIb supernova well past maximum light.
Matheson et al. (\cite{matheson01}) pointed out that the spectrum was almost identical
to one of SN~1993J obtained on day 93 after explosion Matheson et al. (\cite{matheson00}).
Since SN~1993J was a strong radio emitter, it was natural to expect SN~2001gd
to be also strong in radio.
Stockdale et al. (\cite{stockdale02}) detected SN~2001gd on 8 February 2002
at cm-wavelengths with the Very Large Array (VLA), thus confirming the suggestion
that SN~2001gd should be a strong radio emitter.
Their continuous monitoring of SN~2001gd since its first radio detection
has confirmed that SN~2001gd is very similar in
its radio properties to SN~1993J (Stockdale et al.~\cite{stockdale03}),
as suggested by their optical similarity.

We observed SN~2001gd on 26 June 2002 and 8 April 2003 at a frequency of 8.4 GHz,
using an earth-wide global VLBI array (P\'erez-Torres et al.~\cite{mapt04}).
Unfortunately, we were unable to resolve the fine radio structure of SN~2001gd
(Fig.~\ref{fig,sn2001gd}), despite the milliarcsecond resolution of our VLBI array.
Therefore, we had to resort to non-imaging data analysis
to make fits to the visibility data and estimate
the true supernova size at each epoch.
We assumed spherical symmetry for our model fits, and
used the following models:
(i) an optically thin sphere;
(ii) an uniformly bright, circular disk; and
(iii) an optically thin shell of width 25\% the inner radius.
Formally, all three models gave equally good fits to the ($u,v$) data,
so in principle we could not rule out any of them.
Although we suggest the uniformly bright, optically thick disk might be ruled out
as a physical model in 2003.27, since the supernova was optically
thin at 8.4\,GHz, we have left it for comparison purposes. 
The systematic increase in angular diameter in all three models
indicates that the supernova is expanding.

At a distance of 13~Mpc, 100\,$\mu$as correspond to a linear distance of
2$\EE{16}$\,cm.
The average expansion velocity of the supernova can then be written as
22600\,$D_{13}\,\theta\,t^{-1}$\kms, where
$D$=$13\,D_{13}$\,Mpc is the apparent distance to the supernova,
$\theta$ the angular radius in $\mu$as,
and $t$ the time since explosion, in days.
Hence, the average expansion speed of SN~2001gd on 26 June 2002 ($t=296^{\rm d}$),
ranges from $v_{\rm exp}\,=\,12700\pm$2600\kms\ (optically thin shell), up to
14900$\pm$3000\kms\ (optically thin sphere).
For our second epoch ($t=583^{\rm d}$), we obtain velocities
between $7200\pm$700\kms\ (shell) and 8600$\pm$1800\kms\ (sphere).
For any given model, the expansion speeds seem to decrease, but this decrement
is only marginal, so the expansion is compatible with a non-decelerated ejecta.
Thus, future VLBI observations are necessary to confirm, or reject, the validity
of the deceleration.

We can estimate a minimum total energy and a minimum magnetic field
for SN~2001gd, as we have done previously for SN~1986J.
From our VLA observations on 8 April 2003, we determined an spectral index
of $\alpha =-1.05 \pm 0.08$
($S_\nu \propto \nu^{\alpha}$) between 1.4 and 43\,GHz, and
$S_{8.4 \rm GHz} = 1.02\pm0.02$. With those values, we obtain
$L_\nu\,=\,(6.0\pm0.1) \times 10^{36}\,D_{13}^2$\,\ergs
between 1.4 and 43\,GHz.
As $R$, we take the linear size corresponding to half the angular size of SN~2001gd
measured in the 8 April 2003 epoch, $\theta = (190\pm40)\mu$as, which translates into
to a linear size of $R = D\,\cdot\,\theta=(3.8\pm0.8)\EE{16}$\,cm.
With these values, we get

$ E_{\rm min}\,\approx\,1.6\EE{46} \,(1 + k)^{4/7}\, 
   \phi_{0.5}^{3/7}\,
   \theta_{190}^{9/7}\,   
   D_{13}^{17/7}\,  {\rm erg}
$

$
 B_{\rm min}\,\approx\, 43\,(1 + k)^{2/7}\, 
   \phi_{0.5}^{-2/7}\, 
   \theta_{190}^{-6/7}\,   
   D_{13}^{-2/7}  {\rm mG}
$

Since $1 \leq k \leq 2000$,
$E_{\rm min}$ is in the range  $(2.4\EE{46}$--$1.3\EE{48})$\,erg,
and the equipartition magnetic field in the range (52--376)\,mG.
Thus, for a typical energy of 10$^{51}$\,erg for the explosion of SN~2001gd, the fraction
of energy necessary to power the radio emission of SN~2001gd is quite modest.
The upper range of the equipartition magnetic field is in marginal agreement with
that obtained for SN~1993J at the same radius
($B\approx$\,1.8\,G, \cite{fb98};
$B\approx$\,0.5\,G, \cite{mapt01}).
We note that our inferred equipartition magnetic field 
cannot be explained by compression 
of the pre-existing field in the circumstellar wind.
Following the same reasoning made for SN~1986J, we obtain that 
$ B_{\rm csm} \lsim (\mdot\,v_w)^{1/2}\,r^{-1}\,\approx\, 8\,(\mdot_{-4}\,v_{10})^{1/2}\, r_{16}^{-1}\,{\rm mG} 
$
For $r=3.8\,\EE{16}$\,cm and a mass loss rate similar to that of 
SN~1993J ($\mdot = 5\EE{-5}\,\msunyr$),
we obtain $B_{\rm csm} \lsim 1.5$\,mG, which is a factor about 
35 to 250 times smaller than the equipartition field.
Thus, it seems that to explain the radio emission from SN~2001gd, 
we also need to invoke amplification mechanisms other than compression of 
the circumstellar magnetic field, e.g., turbulent amplification.

\section{Summary}

We have presented results from VLBI observations of two Type II supernovae, 
SN~1986J and SN~2001gd. Those results are 
summarized in Table 1, where we have also included for comparison 
results obtained from 
radio observations on two other Type II supernovae, SN~1979C and SN~1993J.

VLBI observations of Type II supernovae suggest that if particles and fields are 
not far from equipartition,  then it seems clear that amplification mechanisms 
other than compression from the
existing circumstellar magnetic field must be acting, in order to 
explain the observed synchrotron radio emission. 
Turbulent amplification seems to be the most promising mechanism.

We would like to end this contribution by stressing that very-long-baseline 
interferometry (VLBI) observations of radio supernovae are a powerful tool to 
probe the circumstellar interaction that takes place after a supernova explodes.
We recall the VLBI community that we have had to wait 11 years to have a really
''SN1993J''-like event, and even so at a distance four times greater. 
Since those supernovae belong to the rare group of
radio supernovae that permit their follow-up with VLBI, we cannot miss the few
opportunities we have of observing and monitoring them.
Unexpected results, like the discovery of a pulsar nebula or the accretion
of material onto a black hole, might be
waiting for us to unveil them, sometimes many years after 
the supernova explosion
(e.g. Bietenholz, Bartel \& Rupen~\cite{bbr04}).

\begin{acknowledgements}
MAPT acknowledges support of the Spanish National programme Ram\'on y Cajal. 
The European VLBI Network is a joint facility of European, Chinese, 
South African and other radio astronomy institutes funded by their 
national research councils.
\end{acknowledgements}

\end{document}